\documentclass[a4paper,11pt]{article}
\usepackage{pos}

\title{Probing CP Violation in Neutrino-antineutrino Oscillations with Non-unitary Flavor Mixing}

\author*[a,b]{Yilin Wang}
\author[a,b]{Shun Zhou}

\affiliation[a]{Institute of High Energy Physics, Chinese Academy of Sciences, Beijing 100049, China}

\affiliation[b]{School of Physical Sciences, University of Chinese Academy of Sciences, Beijing 100049, China}

\emailAdd{wangyilin@ihep.ac.cn}
\emailAdd{zhoush@ihep.ac.cn}

\abstract{If massive neutrinos are Majorana particles, then the lepton number should be violated and neutrino-antineutrino oscillations will take place. In this talk, we present the properties of CP violation in neutrino-antineutrino oscillations with a non-unitary leptonic flavor mixing matrix, which naturally arises in the type-I seesaw model due to the mixing between light and heavy Majorana neutrinos. Taking into account current experimental bounds on the leptonic unitarity violation, we show that the CP asymmetries induced by the non-unitary mixing parameters can significantly deviate from those in the unitarity limit.}

\FullConference{%
  41st International Conference on High Energy physics - ICHEP2022\\
  6-13 July, 2022\\
  Bologna, Italy
}

\begin{document}
\maketitle
\section{Introduction}
Neutrino oscillation experiments have provided us with robust evidence that neutrinos are massive and lepton flavors are mixed~\cite{Zyla:2020zbs}. In order to accommodate tiny neutrino masses, one can naturally extend the Standard Model (SM) by introducing three right-handed neutrino singlets as in the type-I seesaw model~\cite{Minkowski}. Due to the mixing between light and heavy Majorana neutrinos, the flavor mixing matrix $V$ appearing in the charged-current interaction of ordinary light neutrinos becomes non-unitary. In this talk, we concentrate on the CP violation induced by the non-unitary flavor mixing matrix $V$ in the neutrino-antineutrino oscillations. 

The motivation for such an investigation is two-fold. First, since it was suggested by Pontecorvo in 1957~\cite{Pontecorvo:1957cp} that neutrino-antineutrino conversions might occur, there has been great progress in understanding the basic properties of massive neutrinos. Neutrino-antineutrino oscillations indicate lepton number violation, and thus take place only if massive neutrinos are Majorana particles~\cite{Majorana:1937vz}. Apart from neutrinoless double-beta decays ($0\nu \beta \beta$)~\cite{Furry:1939qr}, neutrino-antineutrino oscillations may play an important role in the full determination of two Majorana-type CP-violating phases. Therefore, it is interesting to examine neutrino-antineutrino oscillations in the type-I seesaw model where neutrinos are indeed Majorana particles and the flavor mixing matrix $V$ is intrinsically non-unitary. Second, although neutrino-antineutrino oscillations with a unitary flavor mixing matrix have been studied extensively~\cite{Xing:2013ty}, it is still unclear how different the CP violation with a non-unitary mixing matrix is from that with a unitary one, and how large the deviations can be in light of the latest experimental bounds on unitarity violation. 

\section{Non-unitary Flavor Mixing}

After the spontaneous gauge symmetry breaking, the $6 \times 6$ neutrino mass matrix in the type-I seesaw model can be diagonalized by a unitary matrix as follows
\begin{eqnarray}
	\left( \begin{matrix} V & R \cr S & U \end{matrix} \right)^\dagger \left( \begin{matrix} {\bf 0} & M^{}_{\rm D} \cr M^{\rm T}_{\rm D} & M^{}_{\rm R} \end{matrix} \right) \left( \begin{matrix} V & R \cr S & U \end{matrix} \right)^* = \left( \begin{matrix} \widehat{M}^{}_\nu & {\bf 0} \cr {\bf 0} & \widehat{M}^{}_{\rm R} \end{matrix} \right) \; ,
	\label{eq:unitary}
\end{eqnarray}
where $\widehat{M}^{}_\nu \equiv {\rm Diag}\{m^{}_1, m^{}_2, m^{}_3\}$ and $\widehat{M}^{}_{\rm R} \equiv {\rm Diag}\{M^{}_1, M^{}_2, M^{}_3\}$ with $m^{}_i$ and $M^{}_i$ (for $i = 1, 2, 3$) being the masses of three light and heavy Majorana neutrinos, respectively. Obviously, all the four $3\times 3$ matrices $V$, $R$, $S$ and $U$ themselves are not unitary but satisfy the unitarity conditions. As is well known, the $3 \times 3$ non-unitary mixing matrix $V$ can generally be decomposed into the product of a Hermitian matrix and a unitary matrix, or a lower-triangular matrix and a unitary matrix, namely,
\begin{eqnarray}
	V = \begin{pmatrix}
		1 - \eta_{ee}^{} && -\eta_{e \mu}^{} && -\eta_{e \tau}^{} \\
		-\eta_{e \mu}^\ast && 1 - \eta_{\mu \mu}^{} && -\eta_{\mu \tau}^{} \\
		-\eta_{e \tau}^\ast && -\eta_{\mu \tau}^\ast && 1 - \eta_{\tau \tau}^{}
	\end{pmatrix} \cdot V^\prime = \left( \begin{matrix} \alpha^{}_{11} & 0 & 0 \cr \alpha^{}_{21} & \alpha^{}_{22} & 0 \cr \alpha^{}_{31} & \alpha^{}_{32} & \alpha^{}_{33} \end{matrix} \right) \cdot \widetilde{V} \;.
\label{eq:decomp}
\end{eqnarray}
It is worthwhile to stress that the unitary matrices $V^\prime$ and $\widetilde{V}$ in Eq.~(\ref{eq:decomp}) must be different, since the associated Hermitian matrix $({\bf 1} - \eta)$ and the lower-triangular matrix cannot be exactly identical. The connection between them can be established by performing the QR factorization of the Hermitian matrix $({\bf 1} - \eta)$. As summarized in Appendix A of Ref.~\cite{Wang:2021rsi}, we can obtain
\begin{eqnarray}
	{\bf 1} - \eta \approx \begin{pmatrix}
		1-\eta_{ee}^{} && 0 && 0 \\
		-2 \eta_{e\mu}^* && 1-\eta_{\mu\mu}^{} && 0 \\
		-2 \eta_{e\tau}^* && -2 \eta_{\mu\tau}^* && 1-\eta_{\tau\tau}^{}
	\end{pmatrix} \cdot \begin{pmatrix}
		1 && -\eta_{e\mu}^{} && -\eta_{e\tau}^{} \\
		+\eta_{e\mu}^\ast && 1 && -\eta_{\mu\tau}^{}\\
		+\eta_{e\tau}^\ast && +\eta_{\mu\tau}^\ast && 1
	\end{pmatrix} \;.
	\label{eq:RQ}
\end{eqnarray}

The complex elements of the non-unitary matrix $V$ could bring in extra sources of CP violation, which can be probed in future long-baseline accelerator neutrino oscillation experiments.

\section{CP Asymmetries in Neutrino-antineutrino Oscillations}

Since the flavor mixing matrix is non-unitary, it is more convenient to introduce three neutrino flavor eigenstates (i.e., $|\nu^{}_\alpha\rangle$ for $\alpha = e, \mu, \tau$) as
\begin{eqnarray}
	|\nu_\alpha^{}\rangle = \frac{1}{\sqrt{\left(VV^\dagger_{}\right)_{\alpha\alpha}^{}}}\sum_i V_{\alpha i}^\ast |\nu_i\rangle \; ,
	\label{eq:nualpha}
\end{eqnarray}
which have been properly normalized. The neutrino-antineutrino oscillation amplitudes are usually suppressed by the small ratios $m^{}_i/E$ (for $i = 1, 2, 3$) due to the helicity mismatch between neutrinos and antineutrinos. The CP asymmetries for neutrino-antineutrino oscillations with a non-unitary mixing matrix are found to be
\begin{eqnarray}
	\mathcal{A}_{\alpha \beta}^{} \equiv \frac{P\left(\nu_\alpha^{} \to \overline{\nu}_\beta^{}\right) - P\left(\overline{\nu}_\alpha^{} \to \nu_\beta^{}\right)}{P\left(\nu_\alpha^{} \to \overline{\nu}_\beta^{}\right) + P\left(\overline{\nu}_\alpha^{} \to \nu_\beta^{}\right)} = \frac{\displaystyle 2\sum_{i<j} m_i^{} m_j^{} \mathcal{V}_{\alpha \beta}^{ij} \sin F_{ji}^{}}{\displaystyle \left|\langle m \rangle_{\alpha\beta}^{}\right|^2_{} - 4 \sum_{i<j} m_i^{} m_j^{} \mathcal{C}_{\alpha \beta}^{ij} \sin^2_{} \frac{F_{ji}^{}}{2}} \; ,
	\label{ACPnonuni}
\end{eqnarray}	
where the coefficients $\mathcal{C}_{\alpha\beta}^{ij} \equiv {\rm Re}\left[V_{\alpha i}^{} V_{\beta i}^{} V_{\alpha j}^\ast V_{\beta j}^\ast\right]$, $\mathcal{V}_{\alpha\beta}^{ij} \equiv {\rm Im}\left[V_{\alpha i}^{} V_{\beta i}^{} V_{\alpha j}^\ast V_{\beta j}^\ast\right]$, and the effective neutrino masses $\langle m \rangle_{\alpha\beta}^{} \equiv V_{\alpha 1}^{} V_{\beta 1}^{} m_1^{} + V_{\alpha 2}^{} V_{\beta 2}^{} m_2^{} + V_{\alpha 3}^{} V_{\beta 3}^{} m_3^{}$ (for $\alpha, \beta = e, \mu, \tau$) have been defined. In addition, $F^{}_{ji} \equiv \Delta m^2_{ji} L/(2E)$ for $ji = 21, 31, 32$ stand for the oscillation phases with $L$ being the baseline length and $E$ the neutrino beam energy. 

For illustration, we define the working observables $\varepsilon_{\alpha\beta}^{} \equiv (\mathcal{A}_{\alpha \beta}^{} - \widetilde{\mathcal{A}}_{\alpha \beta}^{}) \times 100\% / \widetilde{\mathcal{A}}_{\alpha \beta}^{}$, where the CP asymmetries in the unitarity limit are denoted by $\widetilde{\mathcal{A}}_{\alpha \beta}^{}$. In minimal seesaw model, where the lightest neutrino is massless and only one Majorana CP-violating phase $\sigma$ exists, we find that $\varepsilon_{\mu\tau}^{}$ can be as large as $206.8 \%$ in the case of normal mass ordering, for $\delta = 195^\circ$, $\sigma = 0^\circ$ and other input parameters from Ref.~\cite{Fernandez-Martinez:2016lgt}. A particularly interesting scenario is to assume all the CP-violating phases in $\widetilde{V}$ in Eq.~(\ref{eq:decomp}) to be vanishing. In this scenario, one can clearly observe that nonzero CP asymmetries stem solely from CP-violating phases in the lower-triangular matrix, characterizing leptonic unitarity violation.

\section{Summary}
In this talk, we have examined the CP asymmetries in the neutrino-antineutrino oscillations in the presence of a non-unitary flavor mixing matrix. The motivation for such an examination is as follows. First, neutrino-antineutrino oscillations occur only when massive neutrinos are Majorana particles. Massive neutrinos are indeed of Majorana nature in the type-I seesaw model, where the flavor mixing matrix of three light neutrinos is intrinsically non-unitary. Second, the non-unitarity introduces extra sources of CP violation. 

By using the QR factorization, we establish the exact connection between the Hermitian and triangular parametrizations of a non-unitary mixing matrix. Then, the CP asymmetries in neutrino-antineutrino oscillations with a non-unitary mixing matrix are calculated. Finally, implementing the latest experimental constraints on the leptonic unitarity violation, we numerically compute the CP asymmetries in the minimal seesaw model.

Although neutrino-antineutrino oscillation probabilities are highly suppressed by $m^2_i/E^2$ for ordinary light neutrinos, our discussions can be generalized to the case of heavy Majorana neutrinos. As heavy Majorana neutrinos of masses around the electroweak scale could be directly produced in the CERN Large Hadron Collider, it is interesting to probe CP asymmetries from heavy neutrino-antineutrino oscillations~\cite{Antusch:2020pnn} and their resonant enhancement if heavy Majorana neutrinos are nearly degenerate in mass~\cite{Bray:2007ru}.

\section*{Acknowledgments}

This work was supported in part by the National Natural Science Foundation of China under grant No.~11835013.

\end{document}